\documentclass[11pt,a4paper,print]{article}

\pdfoutput=1

\usepackage{color}
\usepackage{times,amssymb, amsmath}
\usepackage{graphicx}

\author{
{M.~Petri\c{s}\thanks{mpetris@nipne.ro}, D.~Barto\c{s}, G.~Caragheorgheopol, M.~Petrovici,
L.~R\v{a}dulescu, V.~Simion}\\
\small{"Horia Hulubei" National Institute of Physics and Nuclear Engineering (IFIN-HH)}\\ 
\small{Bucharest, Romania}\\
\\
\textnormal{J.~Fr\"uhauf, M.~Ki\v{s}, P-A.~Loizeau}\\
\small{Gesellschaft f\" ur Schwerionenforschung Darmstadt, Germany}\\
\\
\textnormal{I.~Deppner, N.~Herrmann, C.~Simon}\\
\small{Physikalisches Institut Universit\" at Heidelberg, Heidelberg, Germany}
}

\title{Time and position resolution of high granularity, high counting rate  
MRPC for the inner zone of the CBM-TOF wall}

\begin{document}

%\DeclareGraphicsExtensions{.jpg,.pdf,.png,.mps}

\maketitle

%%%%%%%%%%%%%%%%%%%%%%%%%%%%%%%%%%%%%%%%%%%%%%%%
%%%%%%%%%%%%%%%%%%%%%%%%%%%%%%%%%%%%%%%%%%%%%%%%
%%%%%%%%%%%%%%%%%%%%%%%%%%%%%%%%%%%%%%%%%%%%%%%%

\begin{abstract}
Multi-gap RPC prototypes with readout on a multi-strip electrode were developed for the small polar angle region of the CBM-TOF subdetector, the most demanding zone in terms of granularity and counting rate. The prototypes are based on low resistivity  ($\sim$10$^{10}$ $\Omega$cm) glass electrodes for performing in high counting rate environment. The strip width/pitch size was chosen such  to fulfill the impedance matching with the front-end electronics and the granularity requirements of the innermost zone of the CBM-TOF wall. The in-beam tests  using  secondary particles produced in heavy ion collisions on a Pb target at SIS18 - GSI Darmstadt and SPS - CERN were focused on the performance of the prototype in conditions similar to the ones expected at SIS100/FAIR. An efficiency larger than 98\% and a system time resolution in the order of 70~-~80~ps were obtained in high counting rate and high multiplicity environment. 
\end{abstract}

%%%%%%%%%%%%%%%%%%%%%%%%%%%%%%%%%%%%%%%%%%%%%%%%%%%%%%%%%%%%%%%%%%%
%\keywords{Gaseous detectors, Resistive-plate chambers, Timing detectors, High multiplicity and high counting rate}

%\proceeding{XIII$^{\text{th}}$ Workshop on Resistive Plate Chambers and Related Detectors\\
%  22 - 26 February, 2016\\
%  Gent, Belgium}
%%%%%%%%%%%%%%%%%%%%%%%%%%%%%%%%%%%%%%%%%%%%%%%%%%%%%%%%%%%%%%%%%%%%
\newpage

\tableofcontents

\section{Introduction}
\label{sec:intro}
The Compressed Baryonic Matter (CBM) experiment is a fixed target experiment at the 
 future experimental Facility for Antiproton and Ion Research (FAIR) in Darmstadt. A dedicated research program aims to investigate the properties of the  high net-baryon density matter in 
 nucleus-nucleus collision in an energy range between 2~-~14~GeV/u at the SIS100 accelerator.
 Due to the high interaction rates at which the experiment is designed to run, up to 10$^7$~interactions/s, the detectors of the innermost part of the experimental setup will be exposed to high counting rates and high multiplicities environment.

The Time Of Flight (TOF) subsystem, one of the essential detectors of the CBM  experiment, is foreseen to identify charged hadrons in the angular range covered by the Silicon Tracking System (STS) (2.5$^0$-25$^0$). It covers an active area of about 120~m$^2$, with an approximately rectangular shape. A full system time resolution of at least 80~ps is required, including the electronic contributions and the resolution of the time reference system, at an efficiency of at least 95\%.  This performance should be maintained at a counting rate which,
 very close to the beam pipe, exceeds 3$\cdot$10$^4$~particles/(cm$^2$$\cdot$s) \cite{toftdr}. 
The 5~cm$^2$ area of a readout channel (strip) in the most inner zone is defined by the requirement of an occupancy less than 5\%.  

Our activity has been focused on the development of a Multi-Gap Multi-Strip Resistive Plate Counter (MGMSRPC) prototype 
 for high counting rate and multiplicity environment, as it is  anticipated to be implemented in the inner zone of the   CBM-TOF wall~\cite{toftdr, jinst2014}. 
A narrow strip pitch, 2.54~mm (1.1~mm strip width, 1.44~mm interstrip gap) prototype based on low resistivity glass ($\sim$10$^{10}$ $\Omega$cm), was built as a double stack configuration with 2~x~5 gas gaps of 140~$\mu$m each \cite{jinst}. The transmission line impedance of this prototype has 100~$\Omega$, matching the input impedance of the differential front-end electronics. Its constructive details and very good performance in terms of efficiency, time resolution, position resolutions along and across the strips were already reported \cite{jinst}. 
 However, the number of channels required to equipe the most forward polar angles of the CBM TOF 
wall with such type of MSMGRPC  is quite high ($\sim$ 140,000 electronic channels), with direct consequence on the costs. 
A lower granularity  prototype of 7.4~mm strip pitch (5.8~mm strip width, 1.6~mm interstrip gap), with the same inner architecture, was designed and built further on. 
Due to the larger strip width and double stack configuration, the differential transmission line defined by corresponding strips of the readout electrodes has 50~$\Omega$ impedance. Therefore, an impedance matching with input impedance of fast amplifiers 
%\cite{anghi} 
of 100~$\Omega$ was done at the level of FEE motherboards.
This prototype showed excellent performance in terms of time resolution and efficiency up to local counting rates of 
10$^5$~particles/(cm$^2$$\cdot$s) \cite{jinst} in the in-beam test performed at the COSY facility in J\"ulich and up to a exposure of 10$^4$~particles/(cm$^2$$\cdot$s) all over the counter surface \cite{toftdr} at SIS18 accelerator of GSI Darmstadt. Four prototypes with this larger strip pitch were mounted in a staggered geometry, with overlaps along and across the strips, in order to 
 define a basic architecture which assure a continuous coverage active area in a fixed target experiment. 
This basic architecture was later on implemented  in the proposed design of the modules for the inner zone of the CBM-TOF wall based on MGMSRPCs \cite{toftdr}.

\section{100~Ohm MGMSRPC prototype - constructive details}
\label{sec:constr}
   
In order to achieve a direct matching with the 100~$\Omega$ input impedance of the front-end electronics and in the same time to have the higher granularity required by the most inner zone of the CBM-TOF with a resonable number of readout channels, a new MGMSRPC prototype was designed and built. The strip structure of the readout 
%and high voltage (HV)
 electrodes of 4.19~mm pitch (2.16~mm width and 2.03~mm interstrip gap) was decided based on APLAC \cite{aplac} simulations. The 64 strips of 200~mm length define an active area of 268~x~200 mm$^2$.
As previous developed prototypes, this prototype has also a symmetric double-stack structure of 2~x~5 gas gaps with 140~$\mu$m thickness each gap. 
\begin{figure}[htbp]
\centering % \begin{center}/\end{center} takes some additional vertical space
\includegraphics[width=.61\textwidth, origin=c,angle=0]{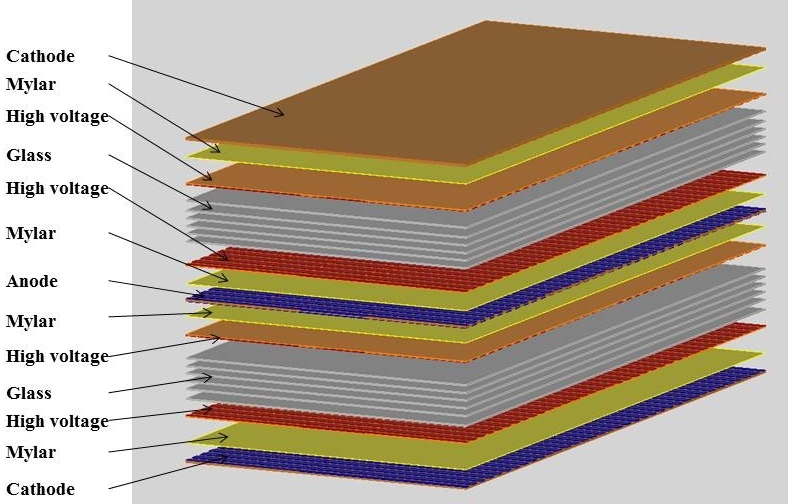}
%\qquad
\includegraphics[width=.26\textwidth,origin=c,angle=0]{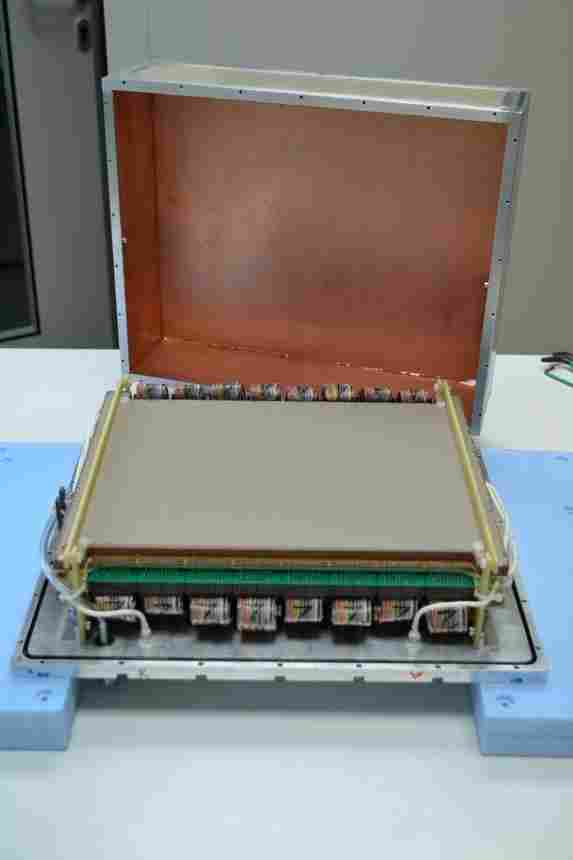}
\caption{\label{fig:f0} Left side: 3D Sketch of the detector configuration. Right side: A photo of the assembled MGMSRPC prototype mounted on the back panel of the closing box. }
\end{figure}
The central read-out electrode (the anode) was made as a single layer Cu strip structure, sandwiched between two thin layers of FR4. The cathodes have strip structure only on the inner side of the FR4 plate. The high voltage (HV) electrodes have the same strip structure as the readout electrodes. The FR4 thickness and two thin mylar layers isolate the readout strips of both anode and cathodes by the strips of HV electrodes of both polarities. The strips of the HV electrodes are in contact with the outermost glass plates of each stack. A 3D sketch of the detector layout is shown in Fig.~\ref{fig:f0}~-~left side. 
  A photo of the assembled prototype mounted on the back pannel of the housing box, called further RPC2013, is presented in Fig.\ref{fig:f0}~-~right-side. The aluminum back pannel of 12~mm thickness holds the mechanical support structure of the counter.
All signals are transported through this back pannel to the front-end electronics (FEE) connected on the outer side of the back pannel. The signals were transfered  via connectors soldered on both sides of a FR4 plate glued on the rectangular opening milled into the back pannel. 

\section{In-beam tests}
\label{sec:beam}
\subsection{Experimental details }
\label{subsec:exp}
\begin{figure}[htbp]
\centering % \begin{center}/\end{center} takes some additional vertical space
\includegraphics[width=.35\textwidth, origin=c,angle=0]{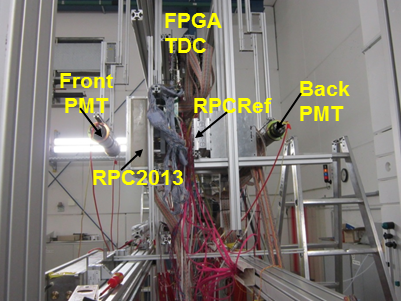}
\includegraphics[width=.47\textwidth, origin=c,angle=0]{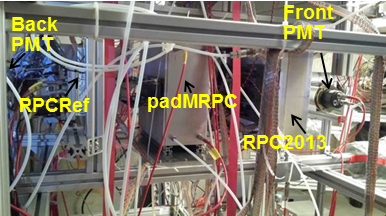}
%\qquad
\caption{\label{fig:f01} Sketch of the experimental setup: left -side at GSI-SIS18 and right - side at CERN-SPS.}
\end{figure}

Two in-beam test campaigns using reaction products produced in heavy ion collisions on a Pb target were focused on testing counter performance in high counting rate and multiplicity environment. The first one, carried out at SIS18 of GSI Darmstadt with a Sm beam of 1.1$\cdot$A~GeV was followed by a CERN SPS beam test  with an Ar beam of 13$\cdot$A~GeV. 
In both cases, the detector under test, RPC2013, was positioned at polar angles of a few degrees relative to the beam axis (at $\sim$7 degrees relative to the beam line in GSI beam test and at $\sim$4 degrees in CERN beam test) in order to get the maximum counting rate accessible  in the experiment. A narrow strip MGMSRPC, called RPCRef, described in \cite{jinst} and positioned behind RPC2013 was used as time reference. Both of them were sandwiched between two plastic scintillators for rate estimates.
In the CERN-SPS experimental set-up between RPC2013 and RPCRef was positioned a MRPC with pad readout from Tsinghua University, China (padMRPC in Fig.\ref{fig:f01} - right side). A diamond detector positioned in front of the target was used as beam reference. The signals delivered by RPC2013 prototype were processed by a new front-end electronics (FEE), called PADI8 \cite{toftdr}, developed within the CBM-TOF collaboration. For RPCRef an older version of PADI, called PADI3 \cite{ciobanu}, was used in the GSI experiment and PADI8 FEE in CERN SPS in-beam test. The digital conversion of the LVDS signals was performed by 32 channel FPGA based TDCs developed at GSI \cite{trb}. 

A gas mixture  of 85\%C$_2$H$_2$F$_4$+ 5\%iso-C$_4$H$_{10}$+10\%SF$_6$ was flushed through the counters at atmospheric pressure in both experiments.
\subsection{Results}
\label{subsec:results}
The first step in the data analysis is the data unpacking including the  calibration and non-linearity corrections of the FPGA TDCs. In the second step an iterative procedure is considered. This includes the calibration of the off-sets in time and position followed by the clusterization which groups the neighboring firing strips, correlated in time and space, in clusters. The mean of the time information delivered at  both ends of a fired strip gives the arrival time of that strip. The cluster time is obtained as the mean of the strip times weighted with the time over threshold information.  
Taking into consideration that in a collision event multiple hits can occur on the counter surface, 
the time of flight (TOF) spectrum was obtained as the time difference of the  best correlated clusters/hits in time and space (the best $\chi$ matching value) between RPC2013 and RPCRef.
In addition to the slewing effect correction, the TOF information was corrected for the spread  of the velocity of the reaction products and hit position. 
\subsubsection{GSI~-~SIS18 Beam Test}
\label{subsubsec:expGSI}
\begin{figure}[htbp]
\centering % \begin{center}/\end{center} takes some additional vertical space
\includegraphics[width=.4\textwidth, origin=c,angle=0]{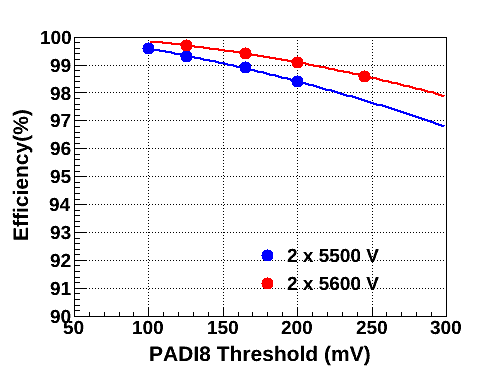}
%\qquad
\includegraphics[width=.4\textwidth,origin=c,angle=0]{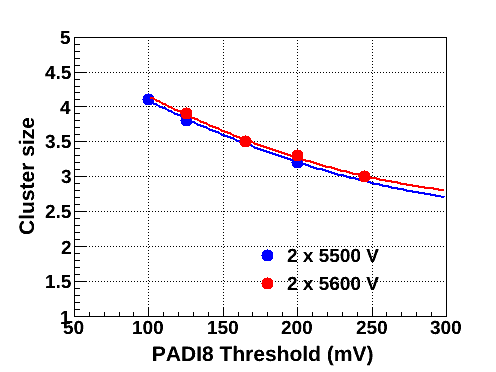}
\caption{\label{fig:f1} RPC2013 efficiency(left side) and cluster size (right side) as a function of PADI8 threshold at two applied HV ($\pm$5.5~kV and $\pm$5.6~kV).}
\end{figure}

The efficiency  obtained with RPC2013 as a function of the PADI8 threshold for two applied high voltages ($\pm$5.5~kV and $\pm$5.6~kV), corresponding to 157~kV/cm and 160~kV/cm electric field respectively, is shown in Fig.~\ref{fig:f1}~-~left side.
The efficiency is calculated as the ratio of the number of matched hits between RPC2013 and RPCRef  and the number of matched hits between RPCRef and diamond detector.
 Even for the largest value of the threshold used in these measurements of 240~mV, the efficiency is larger than 98\%. The statistical errors are within the marker size.

Figure~\ref{fig:f1} - right side presents the mean cluster size as a function of FEE threshold, for the two applied high voltages. The expected behaviour of decreasing of the cluster size  with the increase of the threshold is evidenced. At the highest applied threshold (240~mV) the mean cluster size value is of 3 strips.
As it was already shown in reference \cite{jinst}, this gives the possibility to reconstruct the position across the strips with a position resolution better than the one given by the pitch size divided by the $\sqrt{12}$, in addition to the position information along the strip, obtained from the difference of the times measured at the both ends of the strips. 

\begin{figure}[htbp]
\centering % \begin{center}/\end{center} takes some additional vertical space
\includegraphics[width=.32\textwidth, origin=c,angle=0]{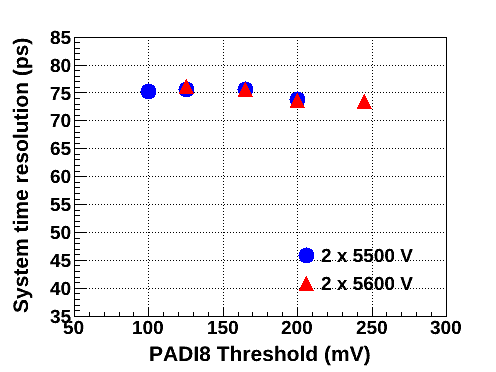}
%\qquad
\includegraphics[width=.32\textwidth, origin=c,angle=0]{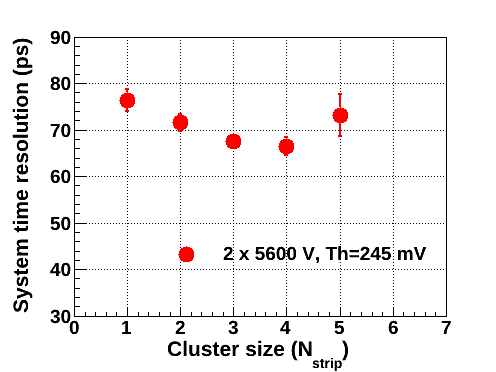}
%\qquad
\includegraphics[width=.32\textwidth, origin=c,angle=0]{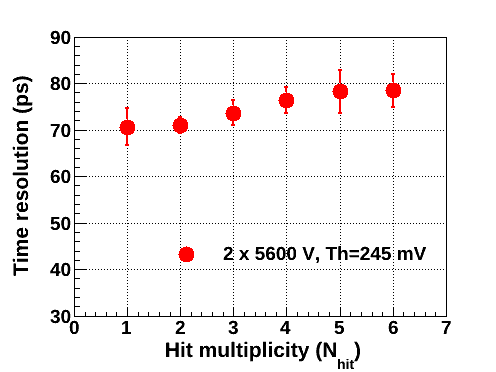}
\caption{\label{fig:f2} System time resolution as a function of: left side - PADI8 threshold at two applied HV ($\pm$5.5~kV and $\pm$5.6~kV); middle side -
 as a function of cluster size, right side - as a function of hit multiplicity, for RPC2013 in GSI in-beam test.}
\end{figure}

 The system time resolution (contributions of both RPC2013 and RPCRef), presented in Fig.~\ref{fig:f2}~-~left side as a function of the PADI8 thresholds, improves slightly with the increasing threshold, reaching a value of 74~ps$\pm$1~ps, including the electronics jitter. 
Figure~\ref{fig:f2}~-~middle presents the system time resolution as a function of cluster size. 
The best time resolution of 67~ps$\pm$2~ps is obtained for events with three or four strips in a cluster, due to the larger amplitude of these signals.
Fig.~\ref{fig:f2} - right side shows the system time resolution as a function of hit multiplicity in RPC2013 prototype, the best time resolution of 71~ps$\pm$4~ps being obtained for single hit events, as it is expected.
The measured particle flux for this in-beam test, estimated using the two platic scintillators shown in Fig.~\ref{fig:f01} - left side, did not exceed 1~kHz/cm$^2$ counting rate.

\subsubsection{CERN-SPS Beam Test}
The same detector prototype was tested at CERN SPS with a lighter ion, Ar, but of higher energy (13$\cdot$A~GeV). 

\begin{figure}[htbp]
\centering % \begin{center}/\end{center} takes some additional vertical space
\includegraphics[width=.45\textwidth, origin=c,angle=0]{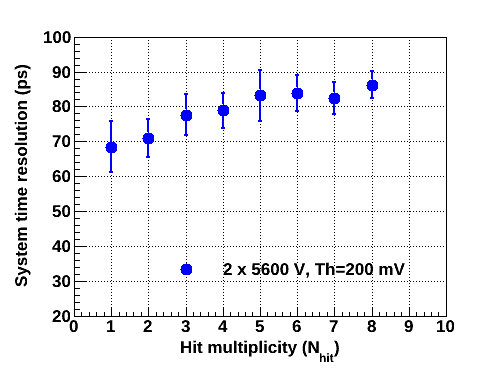}
\includegraphics[width=.45\textwidth, origin=c,angle=0]{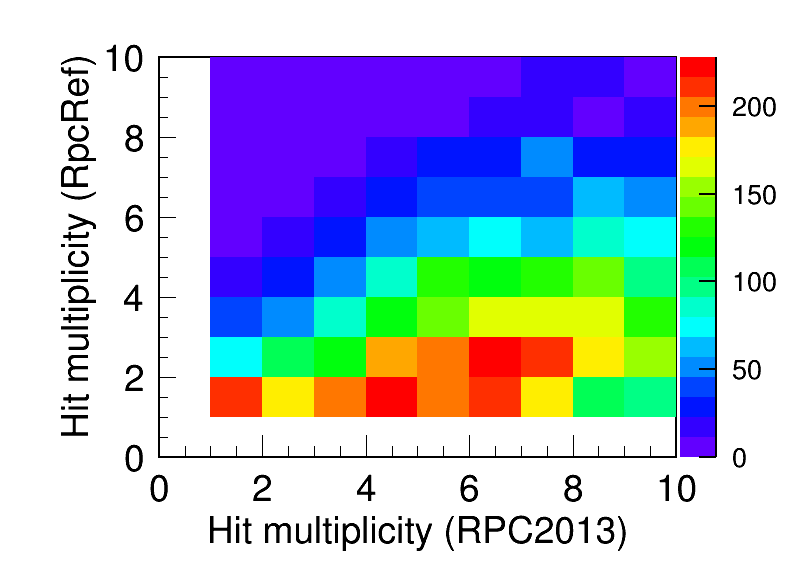}
\caption{\label{fig:f3}System time resolution as a function of hit multiplicity in RPC2013 ($\pm$5.6~kV HV, 200~mV threshold) (left side) and hit multiplicity correlation in RPC2013 and RPCRef (right side) for CERN SPS in-beam test. }
\end{figure}
The system time resolution as a function of hit multiplicity for CERN SPS beam test is presented in Fig.~\ref{fig:f3} - left side. For events with single hit in RPC2013, the obtained time resolution of 69~ps$\pm$7~ps confirms the value obtained in the GSI in-beam test.

The obtained efficiency for RPC2013 is of 99\% and the corresponding mean cluster size 3.6 strips for 5.6~kV applied HV and 200~mV FEE threshold. 
The system time resolution as a function of cluster size for RPC2013 prototype shows the same behaviour as for previous in-beam test, the best obtained time resolution of 75~ps$\pm$4~ps being for clusters of 4 strips. Because in both experiments, (GSI-SIS18 and CERN-SPS), the two counters (RPC2013 and RPCRef) are not exposed to the same conditions (i.e. flux of reaction products, multiplicity - see Fig.~\ref{fig:f3}-right side) in the mentioned geometry of the experiment, they are not exposed to identical conditions. Therefore, only the system time resolution is quoted, the individual contribution of them to the obtained performance being not necessarily the same. 
\begin{figure}[htbp]
\centering % \begin{center}/\end{center} takes some additional vertical space
\includegraphics[width=.45\textwidth, origin=c,angle=0]{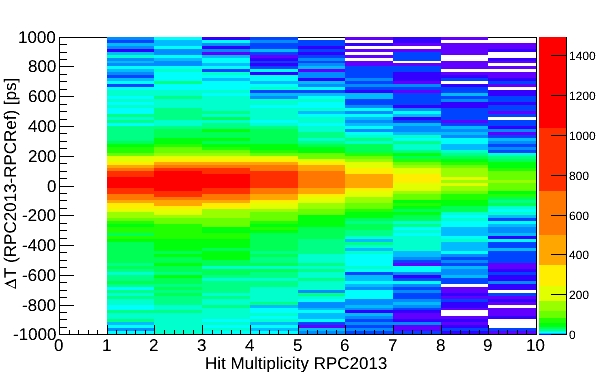}
\includegraphics[width=.4\textwidth, origin=c,angle=0]{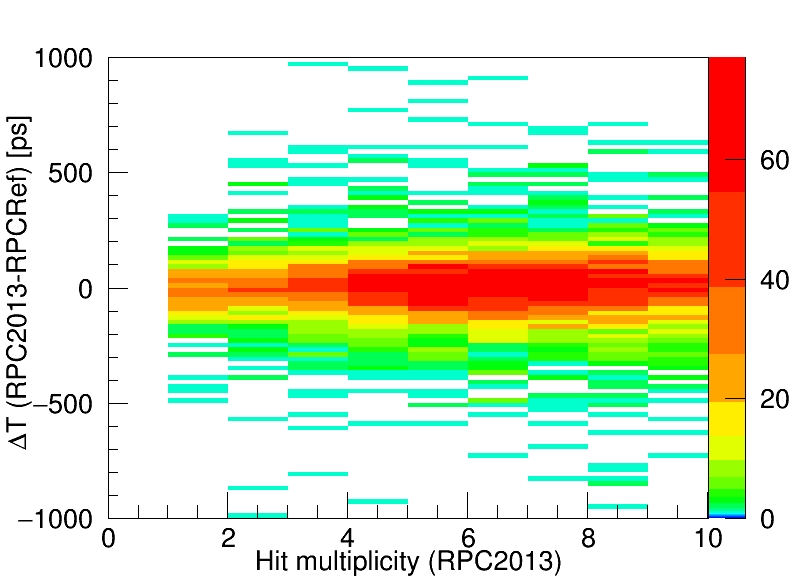}
%\qquad
\caption{\label{fig:f4}Correlation between TOF distribution and hit multiplicity in RPC2013 at 200~mV PADI8 threshold for GSI (left side) and CERN-SPS (right side) in-beam tests. }
\end{figure}

The estimation of the counting rate, using the plastic scintillators positioned in front and behind  of the  experimental setup (see Fig.~\ref{fig:f01}-right side), gives an average value of $\sim$5~kHz/cm$^2$ for the CERN-SPS run used in the present analysis.  
The difference in the hit multiplicity (see Fig.~\ref{fig:f4}) and counting rate between the two mentioned experiments, could explain the slight observed differences in the cluster size and time resolution.

\section{Conclusions}

The performance of a new MGMSRPC prototype is studied in two in-beam tests in conditions close to those anticipated in the operation of the CBM experiment at FAIR. The obtained results for the MGMSRPC prototype operated with the CBM-TOF front-end electronics, of 98\% efficiency with a system time resolution better than 80~ps in high multiplicity and counting rate environment, meet the challenges of the CBM-TOF wall and recommend such an architecture as basic solution for CBM-TOF inner zone.   

%\appendix
%\section{Some title}
%Please always give a title also for appendices.

\section*{Acknowledgements}
%\acknowledgments
We acknowledge V.Aprodu, L.Prodan, A.Radu for their skillful contribution to the construction of detectors.
This work was supported by WP19-HP3 of FP7 - grant agreement 283286, 
contract RO-FAIR F02  and
PN09370103 financed by Romanian National Authority for Scientific Research.


\begin{thebibliography}{9999}
\bibitem{toftdr} CBM-TOF Collaboration, \emph{CBM-TOF Technical Design Report}, (October 2014).
\bibitem{jinst2014}I. Deppner et al., \emph{Journal Of Instrumentation} {\bf 9} (2014) C10014
\bibitem{jinst}  M. Petrovici et al., \emph{Journal Of Instrumentation} {\bf 7} (2012) P11003.
%\bibitem{anghi}
%F. Anghinolfi et al., \emph{Nucl.Instr.and Meth.} {\bf A533} (2004) 183
\bibitem{aplac} http://external.informer.com/aplac.com/
\bibitem{ciobanu} M. Ciobanu et al., \emph{IEEE Trans. Nucl. Sc.} {\bf 61} (2014) 1015.
\bibitem{trb}
\emph{TRB3 Homepage: http://trb.gsi.de/}
\end{thebibliography}
\end{document}